\documentclass{osa-article}

\journal{oe}
\articletype{Research Article}
\begin{document}

\title{Quantifying the x-ray dark-field signal in single-grid imaging}

\author{Ying Ying How,\authormark{1, *} and Kaye Morgan\authormark{1}}

\address{\authormark{1}School of Physics and Astronomy, Monash University, Clayton, VIC, 3800, Australia}

\email{\authormark{*}Ying.How1@monash.edu } %% email address is required



\begin{abstract*}
X-ray dark-field imaging reveals the sample microstructure that is unresolved when using conventional methods of x-ray imaging. In this paper, we derive a new method to extract and quantify the x-ray dark-field signal collected using a single-grid imaging set-up, and relate the signal strength to the number of sample microstructures, \(N\). This was achieved by modelling sample-induced changes to the shadow of the upstream grid, and fitting experimental data to this model. Our results suggested that the dark-field scattering angle from our spherical microstructures is proportional to \(^{2.19}\sqrt{N}\), which deviated from the theoretical model of \(\sqrt{N}\), but was not inconsistent with results from other experimental methods. We believe the approach outlined here can equip quantitative dark-field imaging of small samples, particularly in cases where only one sample exposure is possible, either due to sample movement or radiation dose limitations. Future directions include an extension into directional dark-field imaging. 
\end{abstract*}

%%%%%%%%%%%%%%%%%%%%%%%%%%  body  %%%%%%%%%%%%%%%%%%%%%%%%%%
\section{Introduction}
X-ray imaging has become a widely-adopted technique used to reveal the internal structure of a sample through a non-invasive approach \cite{rontgen}. It has been widely used in clinical practice, material science and security screening \cite{jordan2020handbook}. The image contrast in conventional x-ray imaging originates from differences in the amount of attenuation introduced by each material in the sample. For a sample made up of highly attenuating materials, such as bone or metal, conventional x-ray imaging can provide images of high quality. However, when the sample is made up of weakly attenuating materials such as soft biological tissues or plastics, the image contrast will be degraded significantly. This limitation can be addressed by phase contrast x-ray imaging (PCXI), which is a technique that converts the phase shift experienced by the x-ray wavefield as it passes through the sample into intensity modulations to reveal details that are invisible when using conventional x-ray imaging \cite{wilkins2014}. There are a number of experimental set-ups by which this phase contrast signal can be captured.    

In addition to the attenuation and phase signals, the dark-field signal can also be captured using PCXI set-ups, including analyser-based \cite{rigon2007_abi_3}, grating interferometers \cite{pfeiffer2008}, edge-illumination \cite{endrizzi2014ei,endrizzi2017_3D}, single-grating/speckle-based \cite{wen2010, morgan2013,zanette2014} and propagation-based \cite{gureyev2020}. The dark-field signal originates from small-angle or ultra-small-angle x-ray scattering (SAXS/USAXS) from unresolved sample features and provides a complementary picture of the sample properties, often showing up features that are missing in conventional and phase contrast x-ray images (e.g. \cite{matsunaga2019}). The dark-field signal is useful since it reveals information about sub-micron features in a sample without having to fully resolve those features, which makes it a dose-efficient modality as a detector with large pixels (relative to the sample features) can be used. The dark-field contrast modality can also address the limitation that any measurement of the size of a sample feature will be restricted by the pixel size and the spatial resolution of the imaging system. Therefore, dark-field signal has the potential to play an important role in various fields, especially for medical diagnosis \cite{schleede2012, kitchen2020emphysema,willer2021x}, material science \cite{yang2014,glinz2021} and airport screening \cite{miller2013phase}, with particular benefit seen from the fact it is non-invasive. While dark-field images provide significant value in a qualitative sense, it can also be useful to consider quantifying the dark-field signal to extract more information about the sample.

The dark-field signal has been successfully quantified for several dark-field imaging techniques, including measuring changes in the width of the rocking curve with analyser-based imaging (ABI) \cite{wernick2003_mir,khelashvili2005,rigon2007_abi_3,kitchen2010_scatt} or equivalently in edge-illumination (EI) \cite{Astolfo_2016} and measuring a relative change in the visibility of a reference pattern in grating interferometry (GI) \cite{pfeiffer2008,wang2009_GI_stand,bech2010_coeff,lynch2011_coeff}. The dark-field signal extracted from GI is also related to the sample microstructure size via the linear diffusion coefficient in computed tomography \cite{bech2010_coeff,lynch2011_coeff} and the correlation length \cite{prade2016,harti2017,gkoumas2016} for specific set-ups. Higher-order statistical measures of the scattering distribution, e.g. the kurtosis, can also depend on microstructure size \cite{modregger2017}. The dark-field signal has been related to the sample microstructure properties in a few experimental tests and applications using these techniques \cite{jensen2010,prade2016,gkoumas2016,harti2017,modregger2017, kitchen2020emphysema}. The dark-field signal described by angular spread can be related to the number of scattering interfaces via von Nardroff's equation \cite{von1926refraction}.  Note that when material microstructure is elongated and aligned, resulting in an asymmetrical dark-field signal, this directional dark-field signal has been extracted by capturing the dark-field signal at different orientations of the sample relative to the gratings \cite{jensen2010}.

In this paper, we will focus on single-grid imaging, which is an emerging dark-field imaging technique where the dark-field signal has been extracted qualitatively \cite{bennett2010_invivo, wen2010, morgan2013}, but not yet quantitatively. The main advantage of single-grid imaging is that only a single sample exposure is required during the data acquisition process. The short data acquisition time can help in minimising any motion blur and thus makes this technique feasible for dynamic imaging and time sequence imaging. This technique is closely related to single-exposure speckle-based imaging \cite{berujon2012_sb, morgan2012, zdora2018state}, where the reference grid is simply replaced by a piece of sandpaper to create reference speckle. In this manuscript, we demonstrate the quantification of the x-ray dark-field signal in single-grid imaging and relate the signal to the number of sample microstructures.

We first describe the mathematical modelling of the dark-field signal in single-grid imaging and derive a method for extracting this signal using an explicit cross-correlation approach. This is achieved by modelling the intensity pattern seen behind a grid with a sinusoidal function and looking at how this is reflected in the local cross-correlation between a section of the intensity pattern seen with and without the sample present. We then apply the numerical analysis to synchrotron images, converting visibility loss into an effective scattering angle and relating the angle to the number of microstructures in the sample. Finally, we discuss conditions for quantitative dark-field extraction with a single sample exposure, apply the technique to edge-scattering and to speckle-based images, and look at future directions.

\section{Mathematical model for dark-field effects in single-grid imaging}
This method of analysis is based on a `single-grid' experimental set-up, as shown in Fig. \ref{fig:setup}, where an absorption grid is placed in the illuminating beam, some distance upstream of a detector, and a reference image captured. The sample is then introduced close to the absorption grid, such that with propagation it distorts the grid pattern according to the phase properties of the sample and blurs the grid pattern according to the dark-field properties of the sample \cite{takeda2007,bennett2010_invivo,wen2010, morgan2011_grid, morgan2013}.

\subsection{Modelling the intensity behind a grid} \label{math_inten}
We start by modelling the x-ray intensity seen behind the grid with and without the sample, in one-dimension. Although for an ideal system, we might model the intensity behind a grid as a square wave, we modelled the intensity of the grid as a sinusoidal function here as we observed that the intensity behind the grid is better modelled by a sinusoid in the experimental data. It is worth noting that any regular intensity oscillations, for example, a smoothed square or triangular wave, captured in an experimental setting (finite point spread function and source size) are likely to be well-modelled by a sinusoid, when sampled at around 7-15 pixels per period (typical for this imaging method to balance spatial resolution and range of sensitivity). This is also an appropriate model for the periodic intensity produced by a phase grid \cite{morgan2013, rizzi2013x, gustschin2021}.

We define the intensity seen in the presence of the grid as a function of position $x$ across the image, \(I_{g}(x)\), as a sine function, 
\begin{equation}
    I_{g}(x) = a \sin \left(\frac{2 \pi  x}{p}\right)+b,
\end{equation}
where \(a\) represents the amplitude, \(b\) represents the mean and \(p\) represents the period of the grid. We will refer to \(I_{g}(x)\) as the `grid intensity'. When the sample is introduced, the intensity captured by the detector is shifted transversely and more importantly, has a lower mean gray value and smaller amplitude oscillations due to the presence of sample. This suggests that we can model the `sample-grid intensity', \(I_{sg}(x)\), which refers to the intensity seen in the presence of the grid and sample, with
\begin{equation}\label{Isg}
    I_{sg}(x) = A a \sin\left(\frac{2 \pi  x}{p}\right) + t b,
\end{equation}
where \(A\) represents the change in the amplitude of the grid intensity oscillations that is introduced by the sample and \(t\) represents the transmission of x-ray wavefield passing through the sample, which can have a value ranging from 0 to 1, where 0 means no x-ray wavefield passes through the sample and 1 means the x-ray wavefield is not attenuated. The phase shift of the x-ray wavefield is not taken into consideration here since the sample-grid intensity will effectively be shifted to align with the grid intensity during cross-correlation analysis, as explained later, before we perform the dark-field retrieval. A cross-correlation can be used in single-exposure quantitative single-grid or speckle-based phase retrieval \cite{morgan2012, morgan2013}, to compare the local `grid intensity' with the `sample-grid intensity', and we use the same approach here for dark-field retrieval, for reasons described in the discussion.

The next step is to perform a cross-correlation between \(I_{sg}(x)\) and \(I_{g}(x)\), and then also \(I_{g}(x)\) with itself to allow normalisation of variations in the visibility and local mean intensity of the reference grid image. These two cross-correlations result in \(k(i)\),  

\begin{equation}\label{sgcc}
\begin{split}
    k(i) = (I_{sg} \star I_{g}) (i) &= \int^{p}_{0} \int^{p}_{0} \overline{I_{sg}(x-i)} I_{g}(x) \, dx \,dy \\ 
    k(i) &= \frac{1}{2} A a^2 p^{2} \cos \left(\frac{2 \pi  i}{p}\right)+b^2 p^{2} t,  
\end{split}
\end{equation}
and \(h(i)\),
\begin{equation}\label{ggcc}
    \begin{split}
    h(i) = (I_{g} \star I_{g})(i) &= \int^{p}_{0} \int^{p}_{0} \overline{I_{g}(x-i)} I_{g}(x) \, dx \, dy \\ 
    h(i) &= \frac{1}{2} a^2 p^{2} \cos \left(\frac{2 \pi  i}{p}\right)+b^2 p^{2},
    \end{split}
\end{equation}
respectively. The cross-correlation is performed over one period of grid-induced intensity oscillations instead from \(-\infty\) to \(\infty\) since the sinusoidal function does not converge at infinity, and the cross-correlation is typically performed across one period of the grid within an experimental image.

By comparing the amplitude of \(k(i)\) to \(h(i)\), we can obtain
\begin{equation}\label{A}
\frac{\text{Amplitude(k)}}{\text{Amplitude(h)}} =\frac{\frac{1}{2} A a^2 p^{2}}{\frac{1}{2} a^2 p^{2}} = A,
\end{equation} 
and by comparing the mean of \(k(i)\) to \(h(i)\), we can obtain
\begin{equation}\label{t}
\frac{\text{Mean(k)}}{\text{Mean(h)}} =\frac{b^2 p^{2} t}{b^2 p^{2}} = t .
 \end{equation}

\subsection{Modelling how the dark-field is reflected in the cross-correlation analysis} \label{math_df}
The dark-field signal, \(DF\), extracted from a grating-based technique is defined as the relative change in visibility between the sample-grid intensity (or stepping curve), \(V^{s}\), and the grid-intensity (or stepping curve), \(V^{r}\) \cite{pfeiffer2008},
\begin{equation}\label{DF_signal}
\begin{split}
    DF &= \frac{V^{s}}{V^{r}} = \frac{\frac{I^{s}_{max} - I^{s}_{min}}{I^{s}_{max} + I^{s}_{min}}}{\frac{I^{r}_{max} - I^{r}_{min}}{I^{r}_{max} + I^{r}_{min}}}
    = \frac{\frac{Aa}{tb}}{\frac{a}{b}} 
    = \frac{A}{t},
\end{split}
\end{equation}
where we have used Michelson's definition of visibility, \(V = (I_{max} - I_{min})/(I_{max} + I_{min})\), where \(I_{max}\) is the local maximum intensity and \(I_{min}\) is the local minimum intensity \cite{michelson1995studies}. 

We now have a way to measure \(A\) and \(t\) from the grid-only and sample-grid images (Section \ref{math_inten}), and to use \(A\) and \(t\) to quantify changes in visibility of the reference pattern (Eqn. \ref{DF_signal}). This would be sufficient to collect a visibility-reduction image to describe the dark-field, as typically used in grating interferometry. However we would like to relate the dark-field to quantities that describe the sample, so the next step is to link visibility to a less set-up-specific quantity. We do this by understanding how the set-up parameters such as the propagation distance and grid period relate to the collected visibility-reduction dark-field signal.

\subsection{Modelling how the dark-field signal depends on propagation distance and blur width} \label{math_df_and_width}
The dark-field effect can be modelled as a local blurring of the collected intensity image \cite{paganin2019x, morgan2019applying}, and we can define the blur width geometrically as shown in Fig. \ref{fig:setup}. Since we assume the effective scattering angle, \(\theta\), to be very small, it can be written as,
\begin{equation}\label{scatt_angle}
\begin{split}
    \tan\left(\frac{\theta}{2}\right) \approx \frac{\theta}{2} &= \frac{d/2}{z} \\
    \theta &= \frac{d}{z},
\end{split}
\end{equation}
which can be rearranged into \(d = z \theta\), where \(d\) is the blur width and \(z\) is the distance between the sample and detector (propagation distance). 

We can model the blurring kernel applied to the grid pattern in the presence of sample as a normalised Gaussian function centred at zero, \(G(x)\) \cite[Figure~5]{ kitchen2004}\cite{khromova2004monte,khelashvili2005_gauss}, 
\begin{equation}
    G(x) = \frac{1}{\sqrt{2 \pi } (d/2)} \exp{\left(-\frac{x^2}{2 (d/2)^2}\right)},
\end{equation}
where we have set mean \(\mu = 0\) and we model the blur width, \(d\) as twice the root-mean-squared width of the Gaussian (equivalently the standard deviation), \(\sigma\). The sample-grid intensity can then be modelled by the convolution between the grid intensity and \(G(x)\), 
\begin{equation}\label{IsgSin}
\begin{split}
    I_{sg}(x) = I_{g}(x)* G(x) = a \exp\left(-\frac{2 \pi ^2 (d/2)^2}{p^2}\right) \sin \left(\frac{2 \pi i}{p}\right)+b, 
\end{split}
\end{equation}
where \(i\) is the coordinate in the convolution space. The dark-field signal (defined in Eqn (\ref{DF_signal})) is given by, 
\begin{equation}\label{df_dist_gauss}
    DF = \frac{V^{s}}{V^{r}} = \frac{a \exp\left(-\frac{2 \pi ^2 (d/2)^2}{p^2}\right)}{b} \frac{b}{a} = \exp \left(-\frac{2 \pi ^2 (d/2)^2}{p^2}\right) = \exp \left(-\frac{2 \pi ^{2} z^{2} \theta^{2}}{4p^2}\right),
\end{equation}
where we substituted the definition of scattering width, \(d = z \theta\). Note that a sin or a cos function can be used interchangeably in Eqns. \ref{Isg} and \ref{IsgSin}, since the cross-correlation sinusoid is centred during the analysis process. 

\subsection{Modelling how the dark-field signal relates to the sample properties}
We now have a measure of the dark-field, \(\theta\), which is independent of the set-up used to take that measurement. The next step is to relate that measurement to properties of the sample microstructure. It is possible to model the microstructures as many spheres, in which case the average total deviation angle of an x-ray passing through a number of spheres, \(\alpha_{0}\) is described by \cite{von1926refraction}
\begin{equation}
\label{vonNardroff}
    \alpha_{0} = \left(4\delta^{2}N(\log{\frac{2}{\delta} + 1})\right)^{1/2},
\end{equation}
where \(\delta\) is the complex refractive index decrement of the sphere material and \(N\) is the number of microstructures an x-ray beamlet encounters while propagating along the optical axis. By assuming the sample microstructures, which in our experiment are microspheres, to have the same diameter, we can say that the total thickness of the microspheres, \(T\) is directly proportional to the number of microstructures, \(N\). This implies that the effective scattering angle \(\theta\) is directly proportional to \(\sqrt{T}\) in the case where all the scatterers are of the same size and material.

\section{Experimental details}
We captured an experimental dataset to test this approach to dark-field retrieval and the quantitative nature of the extracted dark-field measure, over a range of propagation distances and hence set-up sensitivities.

\subsection{Experimental setup}
\begin{figure}[hbt!]
\centering\includegraphics[width=14cm]{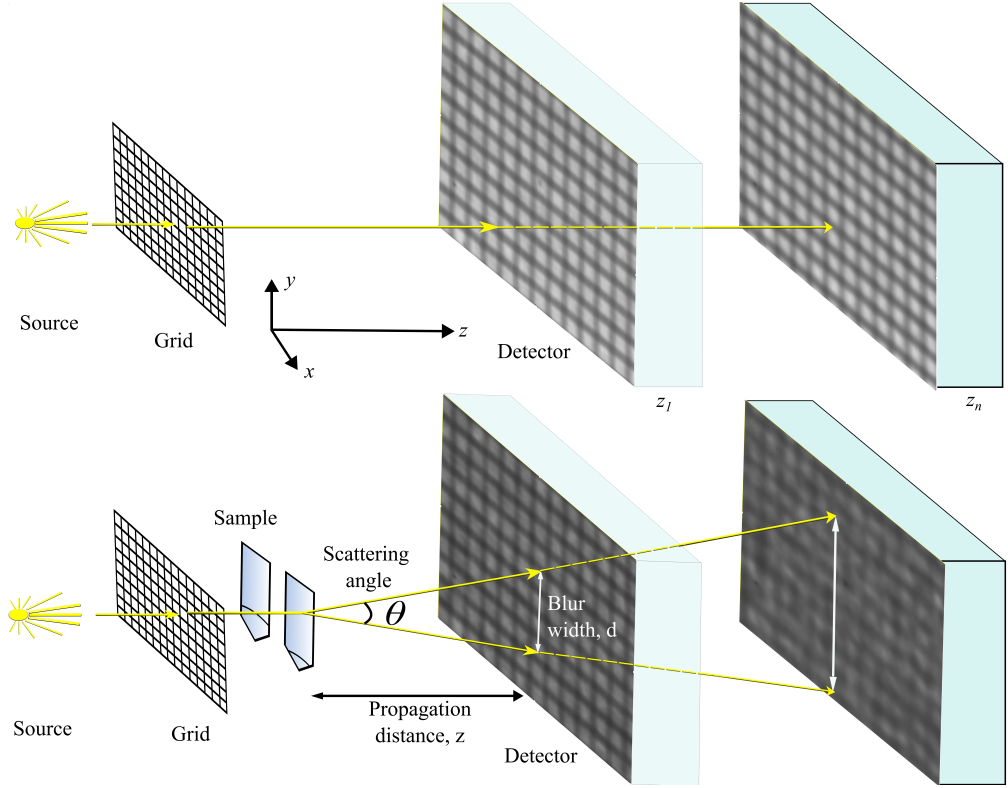}
\caption{Set-up for single-grid imaging with detector being placed at multiple propagation distances, \(z\). The sample can be placed immediately upstream or downstream of the grid. The linear relationship between blur width, \(d\) and \(z\) is described by Eqn. \ref{scatt_angle}. The sample introduces attenuation and significant blurring on the grid pattern formed on the detector especially at larger propagation distance. Note that the scattering of the x-ray has be exaggerated in this figure for visualisation purpose.}
\label{fig:setup}
\end{figure}

The setup of our experiment is the same as the typical single-grid imaging technique, with the addition that we allowed the propagation distance to change as shown in Fig. \ref{fig:setup}. The experiment was performed at the Australian Synchrotron Imaging and Medical Beamline (IMBL). An attenuating grid (a geological stainless steel sieve with holes of 90 \(\mu\)m) was placed 150cm upstream of the sample (as close as the set-up permitted). The sample was allowed to move along the $x$ axis so that it could be moved out of the field of view to capture a grid-only (reference pattern) image. Both the grid and sample were placed together on a table which could be moved to adjust the distance $z$ between the sample and the detector, which was on a separate table. The energy of the x-ray beam used was 30 keV and the exposure time for each image was 650 ms (chosen to fill the dynamic range of the detector). The effective pixel size for the setup was 11.5 \(\mu\)m, using a 25~$\mu$m thick Gadox phosphor coupled to a pco.edge 5.5 sCMOS detector. 

\subsection{Sample preparation}
Our sample began as a PMMA sample tube containing polystyrene microspheres suspended in 10ml of water, with the spheres making up 2.5\% of the volume. Prior to imaging, the samples were spun in a centrifuge for 3 minutes or longer until most of the microspheres precipitated at the bottom of the tubes and formed a solid and an aqueous layer. We then removed the solution using a pipette and left the cap off to allow the liquid to evaporate. Two tubes were imaged, one containing microspheres of 1.0 \(\mu\)m diameter and the other containing microspheres of 8.0 \(\mu\)m diameter. Images were captured at propagation distances ranging from 0.2m to 2.7m with an increment of 0.1m and also at 0.25m to improve sampling at short distance.  At each propagation distance, we took 30 exposures of the `grid intensity' and 30 of the `sample-grid intensity' and averaged each to improve the signal-to-noise ratio.

\section{Numerical analysis/results}
\label{sec:Numerical}
Before analysing the images, we performed a few pre-processing steps including adjusting for temporal changes in flux, adjusting for grid movement and calculating the grid period in pixels. We then performed cross-correlation analyses using the spatial mapping method proposed by Morgan \textit{et al.} \cite{morgan2011_grid}, which determines the distortion of reference pattern directly in real space on an area-by-area basis. We then fitted Eqns. \ref{sgcc} and \ref{ggcc} to the cross-correlation profiles in $x$- and $y$-direction to extract the amplitude and mean of each profile, while fixing \(p\) as the period of the grid and setting $i=0$ as the coordinate of the maximum of the cross-correlation, with sub-pixel accuracy, to account for phase-induced shifts. Examples of this fitting process performed at pixels with and without the presence of sample, along with a histogram to show the confidence of the fitting process are shown in Figs. S2, S3 and S1 respectively in the Supplementary Materials.   

\subsection{Extracting the sample-induced visibility loss}
The next step following the fitting is to extract the change in visibility induced by the sample from the cross-correlation result. It is worth mentioning that this method is designed for isotropic scattering which gives rise to dark-field signal that is not directional. This means that the dark-field signal in each direction is expected to be the same. 

Since our samples are microspheres, we made a reasonable assumption that the blurring of the grid pattern imposed by the sample is symmetric. Following this assumption, we can simplify the 2D cross-correlation result into 1D by focusing on the central row or column of the phase-adjusted cross-correlation when we were trying to fit for the amplitude and mean values. In addition, performing a 2D fit would be more time-consuming and may also be more sensitive to noise. Given that most microstructure is not both elongated and oriented all in the same direction, this approach will work for most samples.

To calculate how the amplitude of the intensity oscillations changes, $A$, we calculate the RMS amplitude in each of the two directions (ie. \(\textrm{Amplitude}(k)=\sqrt{\textrm{Amplitude}(k_x)^2+\textrm{Amplitude}(k_y)^2}\)), and then follow Eqn. \ref{A}. If there is indeed a directional dark-field signal, the RMS measure accurately extracts the typical reduction in visibility. To obtain the transmission value, \(t\), we followed Eqn. \ref{t}, and then simply averaged the transmission from each of the $x$- and $y$-direction fits, since no asymmetry should be present in a scalar value. The transmission \(t\) was corrected in this dataset by an addition of 0.003 to account for the effect of the wide point spread function of the detector, which resulted in transmission value lower than the unity at areas where the sample is not present. This effect should ideally be corrected by performing a deconvolution between the measured point spread function and the raw images, if the measured point spread function is available. The dark-field signal is then calculated by dividing the fitted \(A\) by \(t\), as shown in Eqn \ref{DF_signal}.  

\subsection{Converting visibility measurements to effective scattering angle}
\begin{figure}[htbp]
\centering
\centering\includegraphics[width=\linewidth]{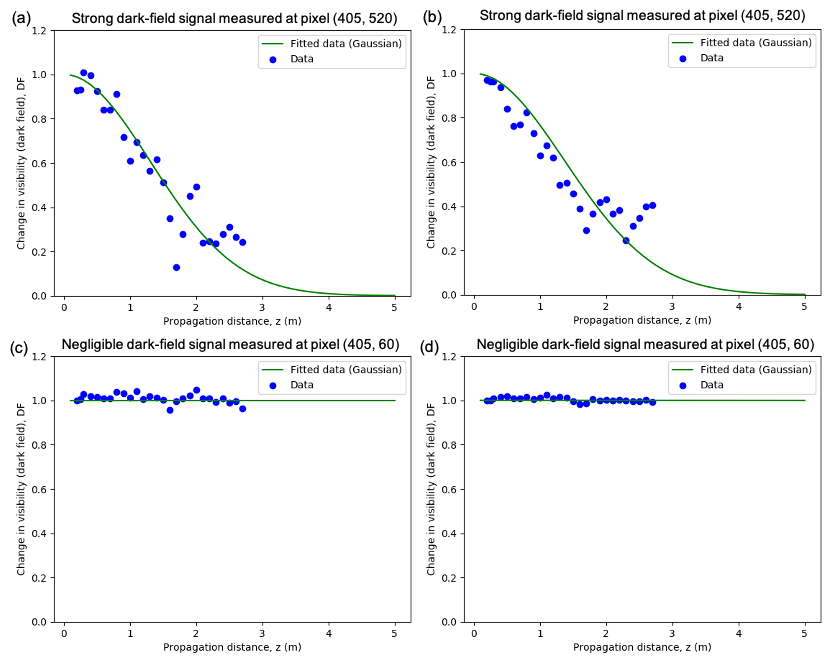}
\caption{A typical example of the dark-field signal measured as a loss in visibility at one pixel from 27 propagation distances (a) \& (c) before and (b) \& (d) after smoothing with a mean square kernel of size 13 pixels. Stronger dark-field signals are measured from pixels in the shadow of the sample ((a) \& (b)) compared to pixels outside the shadow of the sample ((c) \& (d)). The effective scattering angle extracted by fitting Eqn. \ref{df_dist_gauss} to (a) and (b) were \(36.94 \mu\)rad and \(35.25 \mu\)rad, and \(0.0006437 \mu\)rad and \(0.2716 \mu\)rad on (c) and (d).}  
\label{fig:df_plots}
\end{figure}

The images taken at different distances are analysed numerically through the same procedure mentioned in the previous sections to give us a total of 27 dark-field images that each show the change in the visibility of the grid pattern induced by the sample. We then extracted the effective scattering angle by plotting how the dark-field visibility loss at each pixel changes with propagation distance, and fitting to Eqn. \ref{df_dist_gauss}, as shown in Fig. \ref{fig:df_plots}. While we retrieved the dark-field signal pixel-by-pixel, since the dark-field signal measured across the neighbouring pixels within the same period of the grid were expected to be similar, we chose to smooth the dark-field signal images at each distance with an averaging kernel of the same size as the grid period (13 pixels \(\times\) 13 pixels) to reduce the noise in the retrieved dark-field images. The effect of averaging the dark-field image can be seen in Fig. \ref{fig:df_plots}, where the dark-field signals at different distances gave a closer fit after averaging (Fig. \ref{fig:df_plots} (b) \& \ref{fig:df_plots} (d)) compared to the more noisy signals without any averaging (Fig. \ref{fig:df_plots} (a) \& \ref{fig:df_plots} (c)). 

The dark-field visibility signal remained constant at 1.0 when no sample is present (ie. no loss in visibility, see Fig. \ref{fig:df_plots} (c) \& (d)), but decreases as the propagation distance increases with the presence of sample (Fig. \ref{fig:df_plots} (a) \& (b)). This is due to the fact that the blurring width is directly proportional to the propagation distance. 

\subsection{Relating the effective scattering angle to sample properties} 
\label{sec:T}
\begin{figure}[hbt!]
\centering\includegraphics[width=14cm]{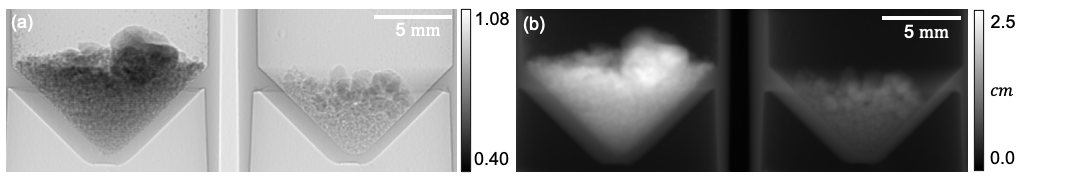}
\caption{(a) Transmission image extracted from single-grid images of two tubes of polystyrene microspheres taken at a propagation distance of 2.7m, after correction for the wide point spread function. (b) Projected thickness recovered from (a) using TIE-based single-image phase retrieval \cite{paganin2002}.}  
\label{fig:trans_and_thick}
\end{figure}

The effective scattering angle is then related to the thickness of the sample, \(T\), which is directly proportional to the number of microstructures, \(N\), since the microstructures in our experiment are all the same size. The thickness of the sample (Fig. \ref{fig:trans_and_thick} (b)) is recovered from the transmission image obtained from images taken at a propagation distance of 2.7m (Fig. \ref{fig:trans_and_thick} (a)) by using the Transport of Intensity Equation (TIE)-based phase retrieval algorithm \cite{paganin2006}, using $\beta = 9.2028 \times 10 ^{-11}$ and $\delta = 2.6286 \times 10 ^{-7} $(polystyrene at 30 keV). This provides a way to test if our results follow the trend described by Eqn. \ref{vonNardroff}, since $N$, the number of microstructures in projection, is proportional to $T$.

\section{Results}\label{df_result} 
\begin{figure}[hbt!]
\centering\includegraphics[width=\linewidth]{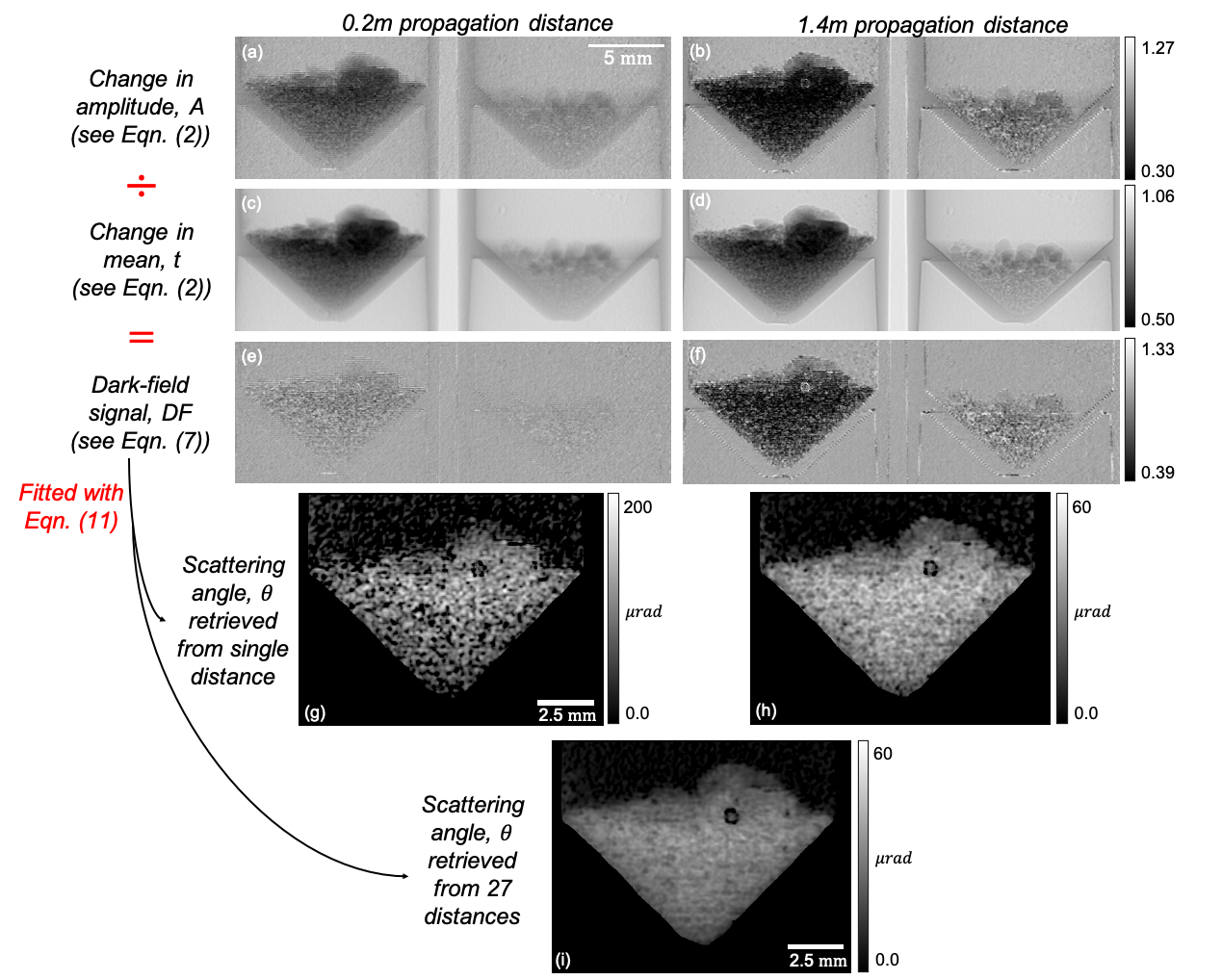}
\caption{Dark-field imaging results obtained from single-grid images of two tubes of polystyrene microspheres, shown here for two different propagation distances - \(0.2\)m and \(1.4\)m. The change in amplitude of intensity oscillations, \(A\) ((a) \& (b)), is divided by the corresponding transmission of x-ray wavefield, \(t\) ((c) \& (d)) to obtain the dark-field signal, \(DF\) ((e) \& (f)). The scattering angles (from the sample tube on the left), \(\theta\) shown in panel (g) and (h) are extracted from the \(DF\) signal in panel (e) and (f) respectively using Eqn. \ref{df_dist_gauss} while \(\theta\) in panel (i) is extracted from the \(DF\) images measured at all 27 distances shown in Fig. \ref{fig:df_plots}. The \(\theta\) image extracted from the larger propagation distance has less noise than the smaller propagation distance, where the blurring effect is weak. The \(\theta\) image retrieved using dark-field signal measured from 27 distances has less noise compared to \(\theta\) image retrieved from single distance.}  
\label{fig:combine}
\end{figure}

We performed the curve fitting analysis on the images taken at 27 different propagation distances to extract the change in amplitude, \(A\) and transmission, \(t\). Examples of \(A\) and \(t\) extracted from images taken at two different propagation distances - \(0.2\)m and \(1.4\)m are shown in Fig. \ref{fig:combine} (a),(b) and (c),(d). The dark-field signal, \(DF\) is then calculated by dividing \(A\) by \(t\) as suggested in Eqn. \ref{DF_signal} and the result is shown in Fig. \ref{fig:combine} (e) and (f). 

\subsection{Effective scattering angle from 27 exposures}
\begin{figure}[hbt!]
\centering\includegraphics[width=10cm]{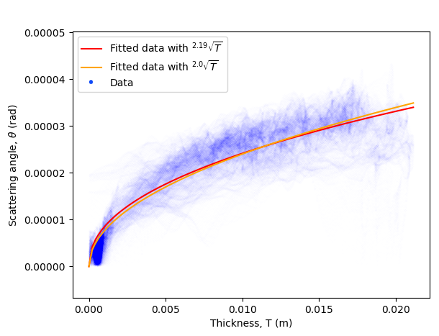}
\caption{The scattering angle, \(\theta\) as a function the sample thickness, \(T\), which is proportional to the number of microstructures $N$. The plot is fitted with \(\theta = m ^{p}\sqrt{T}\), where the exponent of $T$ is allowed to vary, and with \(\theta = m \sqrt{T}\), where \(m\) is a constant, as suggested by von Nardroff \cite{von1926refraction}.} 
\label{fig:theta_thick_2.19}
\end{figure}

To examine the quantitativeness of the extracted dark-field scattering angle, we look at how well this data fits the dark-field scattering model proposed by von Nardroff. Figure \ref{fig:theta_thick_2.19} plots the scattering angle extracted from the dark-field signal measured at 27 distances (as shown in Fig. \ref{fig:combine} (i)) as a function of the sample thickness (see Section \ref{sec:T}), corrected to remove the simulated sample tube thickness. We fitted the data with \(\theta = m \times {T}^{1/p}\), where \(m\) and \(p\) were allowed to vary. From Fig. \ref{fig:theta_thick_2.19}, we can see that the data was best fitted with \(\theta = m \times ^{2.19}\sqrt{T}\) (see Fig. S5 for how the coefficient of determination (\(R^{2}\)) value changes with different p values) instead of \(\theta = m \sqrt{T}\), where \(m\) is a constant, as proposed by von Nardroff  \cite{von1926refraction}. This suggested that the effective scattering angle measured here, \(\theta\), is proportional to \(^{2.19}\sqrt{N}\). 

\begin{figure}[hbt!]
\centering\includegraphics[width=14cm]{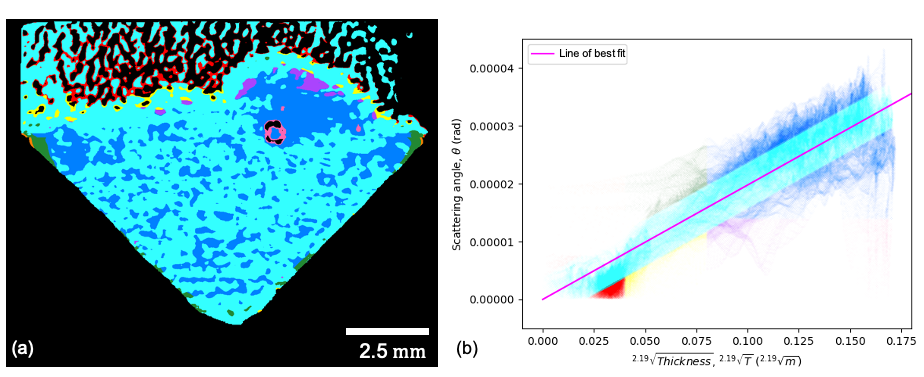}
\caption{Colour-coded mapping of the pixel values within the image of the left tube, showing how well each pixel in (a) the \(\theta\) image retrieved from \(DF\) measured at 27 distances (i.e Fig. \ref{fig:combine}(i)) are described by our model \(\theta = m ^{2.19}\sqrt{T}\). The majority of pixels in (a) is coloured in cyan, indicating that they are well described by the model and sit within the positive and negative \(4\mu\)rad offset from the model. Note that there are a small amount of data points coloured in pink located at the bottom right of the plot in (b). The thickness and scattering angle pixel values are shown for the entirety of the experimental image in Fig. S4.} 
\label{fig:theta_plot_color2_27}
\end{figure}

In Fig. \ref{fig:theta_plot_color2_27}, we aim to show how well each pixel in the image fits the model. The data points that are located within the range of 4 \(\mu\)rad from the line of best fit are labelled in cyan and are distributed evenly across the image. 

\subsection{Effective scattering angle from a single sample exposure}\label{theta_vs_thickness_one} 
\begin{figure}[hbt!]
\centering\includegraphics[width=13cm]{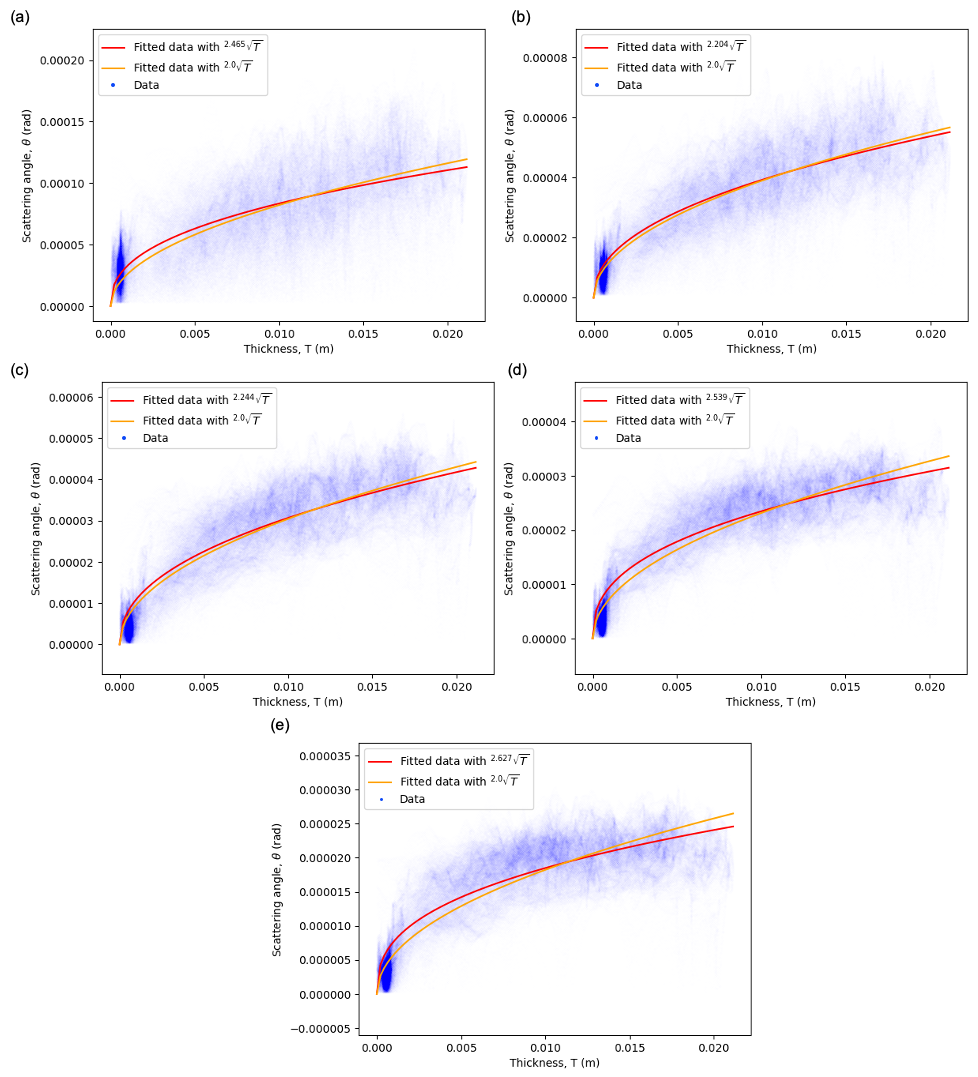}
\caption{The scattering angle, \(\theta\) as a function of the sample thickness, \(T\) (which is proportional to $N$), extracted from dark-field signals measured at images taken at a single distance of (a) 0.2m (b) 0.7m (c) 1.4m (d) 2.1m and (e) 2.7m . The line of best fit is \(\theta \propto ^{p}\sqrt{T}\), where \(p\) corresponds to \(2.456\), \(2.204\), \(2.244\), \(2.539\) and \(2.627\) respectively. The shorter distances show significant noise and the larger distances show saturation for large thicknesses, suggesting that quantitative dark-field imaging with single exposure needs to be performed with a reasonable (not too short or too long) propagation distance to obtain consistent quantitative results.} 
\label{fig:theta_plot_one_res}
\end{figure}

Given the motivation to minimise imaging time and dose and the single-exposure nature of the single-grid technique, it is interesting to examine how quantitative the results are at a single distance instead of using images from multiple distances. The scattering angle, \(\theta\) was extracted from images obtained at propagation distance of 0.2m, 0.7m, 1.4m, 2.1m and 2.7m. Figure \ref{fig:theta_plot_one_res} shows how the scattering angle extracted at each single distance relates to the sample thickness (i.e number of sample microstructures). Examples of the scattering angle extracted at 0.2m and 1.4m are shown in Fig. \ref{fig:combine}(g) and \ref{fig:combine}(h) respectively.  

A significantly wider spread in scattering angle is extracted at shorter distances, such as 0.2m and 0.7m (Fig. \ref{fig:theta_plot_one_res}(a) and (b)). A saturation in the dark-field signal/scattering angle is observed from images taken at larger distances such as 2.1m and 2.7m (Fig. \ref{fig:theta_plot_one_res}(d) and (e)).  

The magnitude of the spread in scattering angle at a given thickness is related to the position of the data points in the plot of \(DF\) against \(z\) (see Fig. \ref{fig:df_plots} (b)). The fit is more robust to noise for data points at larger propagation distances where the blurring effect is strong and easily measured, which results in less uncertainty in the \(\theta\) values extracted. However, if the propagation distance is too large and the grid pattern is so smoothed that it reaches comparable visibility to sample features, then the dark-field signal begins to saturate and hence does not correctly predict the scattering angle for those large thicknesses where the signal is strongest. This effect can be seen in the flattening of the curve in Fig. \ref{fig:theta_plot_one_res} (e).

\section{Further applications of the approach}

While the key aim of this paper was to design an analytical approach to quantitatively retrieve x-ray dark-field from a single-grid set-up, we also want to show how well this approach works in a range of scenarios.

\subsection{Dark-field from edges}
\begin{figure}[htbp]
\centering
\includegraphics[width=\linewidth]{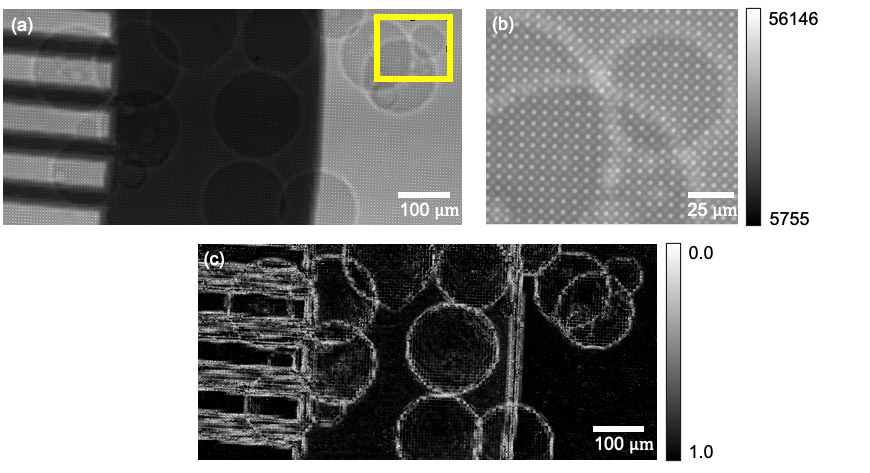}
\caption{(a) Raw image of a phase grid illuminating a gold EM grating and PMMA spheres\cite{groenendijk2020material} (see main text for experimental details). (b) Magnified region from the yellow rectangular area in (a), showing reduced visibility of the grid pattern due to edge scattering. (c) The dark-field signal \(DF\) extracted from (a) using the algorithm presented here.}  
\label{fig:gold_pmma_res}
\end{figure}

Dark-field signal is primarily seen from microstructure, but can also be seen from sample edges. To test the ability of our technique to detect dark-field from edges, we examined some high-resolution single-grid images. Figure \ref{fig:gold_pmma_res} (a) shows a raw image taken using a phase grid to study a gold EM grid and PMMA spheres, also used in  \cite{groenendijk2020material}, using 25 keV x-rays and 0.722 \(\mu\)m effective pixel size. From the magnified image in Fig. \ref{fig:gold_pmma_res} (b), we can see that the grid pattern has lower visibility near the edges of the PMMA spheres due to the edge scattering effect. Figure \ref{fig:gold_pmma_res} (c) shows the dark-field visibility signal, \(DF\), extracted from Fig. \ref{fig:gold_pmma_res} (a) using the algorithm presented in this paper. The blurring of the grid pattern near the edges of the PMMA spheres had been successfully detected as indicated by the stronger signals near the edges of the spheres as shown in Fig. \ref{fig:gold_pmma_res} (c). It is worth mentioning that dark-field from edges was not observed in the dark-field images extracted from the microsphere sample that this manuscript focuses on. We believe this is due to the fact that the dark-field signals produced by our microspheres were much stronger than the edge effect near the edges of the sample tube and thus the edge effect was not visible in our dark-field images. In addition, the pixel size used in the experiment of Fig. \ref{fig:gold_pmma_res} was 0.722 \(\mu\)m \cite{groenendijk2020material}, which was only 6\% of the pixel size used to image the microsphere sample, and thus provided a higher sensitivity to the dark-field signal. 
Our algorithm is therefore able to detect the blurring of grid pattern due to the scattering of the x-ray wavefield from the edges of an object.

\subsection{Application to speckle-based imaging}
\label{sec:speckle}
Theoretically, the approach outlined in this paper could be applied to images captured using any reference pattern. This is because the dark-field signal is extracted from the cross-correlation analyses, which works equally well with any type of reference pattern \cite{morgan2011_grid,morgan2012} and does not require a periodic reference in the way a Fourier technique does. Furthermore, the sine function we use to fit the local intensity is an appropriate model for any local oscillation in intensity. Therefore, we tested our algorithm on a speckle-based dataset \cite{morgan2012, berujon2012_sb, zdora2018state}, acquired at the European Synchrotron Radiation Facility (ESRF), shown in both Berujon \textit{et al.}, 2016 \cite{berujon2016x} and Pavlov \textit{et al.}, 2020 \cite{pavlov2020x}, where sandpaper with grit size P800 was used in the system as a reference pattern instead of an attenuating grid. The sample used in this experiment was a red currant, imaged with 17keV x-rays, and the effective pixel size of the setup was 5.8 \(\mu\)m. The sample was placed 0.5m downstream of the sandpaper and 1m upstream of the detector. Images were taken with the sandpaper at 6 different transverse locations with and without the sample. 

\begin{figure}[htbp]
\centering
\includegraphics[width=\linewidth]{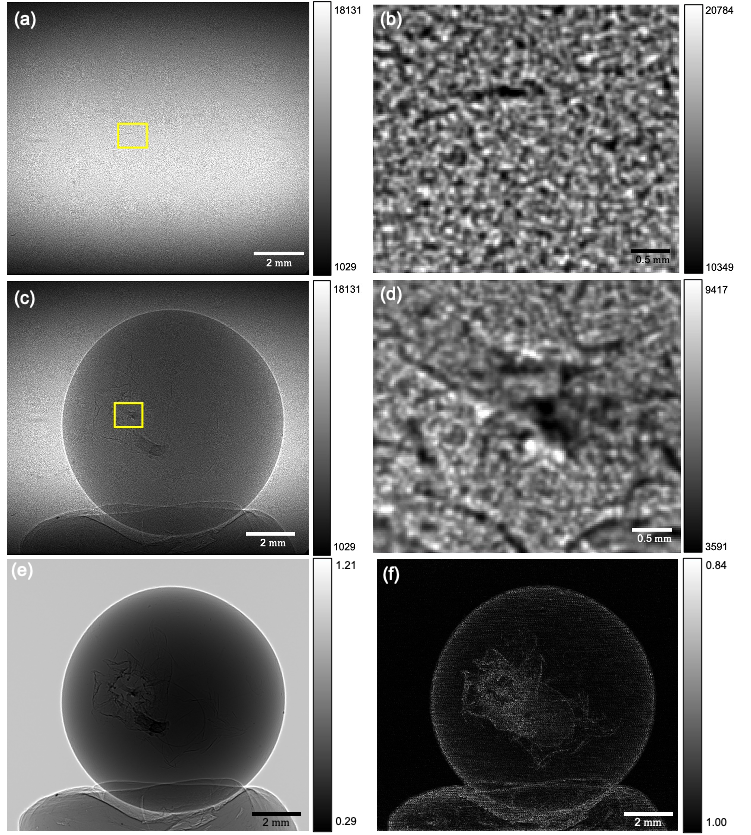}
\caption{The raw (a) speckle-only image and (b) magnified region from the yellow rectangular area in (a) \cite{berujon2016x}. The raw (c) sample-speckle image and (d) magnified region from the yellow rectangular area in (c), where the sample is a red currant (see main text for experimental details). Images of (e) transmission, \(t\) and (f) the dark-field signal, \(DF\) are extracted using the approach described in this manuscript, and taking the median value over 6 images. }
\label{fig:SBI_raw}
\end{figure}

Figure \ref{fig:SBI_raw} (a) and \ref{fig:SBI_raw} (c) shows the raw speckle-only image and sample-speckle image (along with the magnified image of the yellow-rectangular area in the corresponding image) at one position of the sandpaper. In Fig. \ref{fig:SBI_raw} (d), the sample has locally reduced the visibility of the speckle pattern. We performed the analysis described in this manuscript using the speckle-only image and the sample-speckle image with the sandpaper located in the same position. Here, to reduce the noise in the images, we show the median value of each pixel over the 6 images for both the transmission (Fig. \ref{fig:SBI_raw} (e)) and dark-field (Fig. \ref{fig:SBI_raw} (f)) images. We observed dark-field signal near the edges of the sample, inside the currant where the seed/stem sits, and also from the mounting under the currant. The dark-field image also agrees qualitatively with that seen in Pavlov \textit{et al.}, 2020 \cite{pavlov2020x}, where an intrinsic approach is used to capture the dark-field, indicating that our approach is also applicable to speckle-based imaging.  

\section{Discussion} 
In this paper, we have designed and demonstrated a new approach to extract and quantify the dark-field signal using a single-grid imaging set-up. This has been achieved by constructing a mathematical model (Section \ref{math_inten}) and designing a numerical algorithm (Section \ref{sec:Numerical}) to extract quantitative dark-field images. We see qualitative agreement between the results reported in Section \ref{df_result} and dark-field images seen at other set-ups, with signal generated by unresolved microstructure. In addition, we have shown quantitative dark-field signal retrieval from a single-grid PCXI set-up. The recovered dark-field signal is consistent with the model proposed by von Nardroff in that the scattering angle is roughly proportional to the square root of the number of microstructures through which the beam passes. We have successfully applied our algorithm on data obtained using a single-grid imaging setup and on data obtained using speckle-based imaging. Nevertheless, some unexpected behaviour was observed in the dark-field signals we extracted. We have split the remainder of the discussion into a number of sections focused on interesting observations and future directions.

\subsection{Accounting for propagation-based edge effects}\label{TIE_discussion}
Some abnormal dark-field values were observed, especially at larger propagation distances, where a small number of pixels near the edges of the sample tube had visibility greater than 1, which was outside the range of the expected values for the dark-field signal (\(0-1\)). This may be due to the fact that the propagation-based phase contrast effect is generally not taken into consideration in the grating-based imaging techniques, including single-grid imaging. Our model/algorithm has considered the bright fringes near the edges of the sample tubes that comes from the focusing of the x-ray beam as ``negative'' attenuation due to the increase in mean value or amplitude. One possible way to tackle this problem is by removing the propagation-based phase contrast effect from the raw images prior to the dark-field analysis, using an approach similar to that performed by Groenendijk \textit{et al.} \cite{groenendijk2020material}. In the approach taken by Groenendijk \textit{et al.}, the propagation-based phase effect was simulated by propagating a wavefield with uniform intensity numerically through the angular spectrum approach \cite{Nielsen} using phase information retrieved from the images. The simulated propagation-based phase effect was then divided out of the attenuation images in order to perform material decomposition. Another way to solve this issue would be by incorporating the propagation-based phase contrast effects into the retrieval process using the X-ray Fokker-Planck equation \cite{paganin2019x} as suggested by Morgan \& Paganin \cite{morgan2019applying}. 

\subsection{A quantitative dark-field model}
A second interesting point to discuss is the extracted quantitative measures for the scattering angle as a function of the number ($N$)/thickness($T \propto N$) of microstructures of typical thickness $a$. In Section \ref{df_result}, we have shown that the scattering angle measured by our data was best-fitted as proportional to \(^{2.19}\sqrt{T} \propto ^{2.19}\sqrt{N}\) instead of \(\sqrt{N}\) as suggested by von Nardroff \cite{von1926refraction}. This is not inconsistent with the result obtained by Kitchen \textit{et al.} \cite{kitchen2020emphysema}, who used an analyser-crytal approach to measure the scattering angle from PMMA microspheres and found the angle to be directly proportional to \(^{2.2627}\sqrt{N}\). The scattering angles in both cases were proportional to a root with a factor greater than 2 of the number of microstructures in the beam path. We suspected that the deviation of our result from the model proposed by von Nardroff may be due to anomalous diffusion of the scattered x-ray wavefield. To further explore the relationship between the sample microstructures and the scattering angle, we could repeat the analysis on the images obtained with other diameter microstructures. 

\subsection{Optimal single-exposure imaging}\label{single_discuss}
In this experiment, we acquired images at 27 different propagation distances to accurately measure how the scattering angle relates to the number of microstructures in the sample. However, in practice, it is often useful to use a single sample exposure only, to minimise motion blurring due to a moving sample and radiation dose delivered to the sample. A single-exposure approach also allows time-sequence imaging to capture a dynamic process like a biological response to a treatment \cite{morgan2014vivo} (noting that repeated processes can use multiple images \cite{gradl2018}). Therefore, the question then turns to whether the propagation distance between the sample and the grid can affect the accuracy of the results obtained. Generally, it is not recommended to image the sample with the detector being placed immediately downstream of the sample since this will result in a weak blurring effect and it is also not recommended to image the sample with a large propagation distance since this can result in the grid pattern being blurred out too significantly and the cross-correlation analysis may fail. In the second case, we have a saturated dark-field signal where the visibility of the grid falls below the visibility of other features in the image, and sensitivity to a greater dark-field signal is lost.   

In Section \ref{df_result} and \ref{theta_vs_thickness_one}, we showed the images for the scattering angles extracted using dark-field signals measured using one distance only, for 0.2m and 1.4m (see Fig. \ref{fig:combine} (g) and (h)) and the plots showing how these dark-field effective scattering angles measured at a single distance were related to the thickness of the microspheres (see Fig. \ref{fig:theta_plot_one_res}) respectively. At smaller propagation distances, we obtained a wider range of scattering angles for a given thickness compared to larger propagation distances, showing the low sensitivity at these short distances makes measurements less reliable. However, the scattering angle obtained with dark-field signals extracted at shorter distance still varied with sample thickness with a fit that is closely matched with the fit obtained in Fig. \ref{fig:theta_thick_2.19}, despite having a wider range for each sample thickness. For the larger propagation distances, it seems the dark-field is saturated in the sense that we can only just see the grid in the raw images, and the visible intensity variations are largely PBI speckle from the microspheres. In Fig. \ref{fig:theta_plot_one_res} (e), the plot is flattened out for large sample thickness when it should be increasing in dark-field, which may indicate the data is not fitting the model anymore. Figure \ref{fig:df_plots} (b) also shows that the visibility signal does not decrease as expected after around 2m propagation, suggesting a saturated dark-field signal. While Fig. \ref{fig:theta_plot_one_res} (d) and \ref{fig:theta_plot_one_res} (e) show the scattering angle measured is consistent up to 5mm thickness, the scattering angle then `saturates' for larger thicknesses. Although Fig. \ref{fig:combine} (g) and (h) had different ranges of values, they are qualitatively similar. In summary, shorter propagation distance can result in weaker dark-field signal with more variability, while larger propagation distance typically results in a stronger dark-field signal with a smaller variability but measurements will not follow the model laid out here if the visibility signal is saturated. Essentially, we can tune the sensitivity of the system to the dark-field signal by changing the propagation distance. Custom tests would be required to determine the optimum propagation distance for each sample, but we would expect that a suitable propagation distance would show a noticable drop in grid visibility across the sample (e.g. 0.5, see Fig. \ref{fig:df_plots}).

\subsection{Directional dark-field imaging}\label{df_2D}
Our approach to quantifying the dark-field signal could be extended into two dimensions, which would be particularly useful for samples that produce a directional dark-field signal \cite{jensen2010}. The grid-only intensity, \(I_{g}(x,y)\) can be modelled as a two-dimensional sine function and the sample-grid intensity, \(I_{sg}(x,y)\) can be modelled as the convolution between grid-only intensity with a zero-centred asymmetric Gaussian blurring function. A two-dimensional fitting can be performed on the cross-correlation results to extract the magnitude and direction of the blurring contributed by the sample and thus opening up a new area of study on directional dark-field signal. 

\subsection{Advantages and disadvantages of this approach}
The primary advantage of single-grid imaging is that only a single sample exposure is required for the data acquisition process. This makes the approach more feasible for dynamic imaging and time sequence imaging, compared to the other phase contrast x-ray imaging techniques such as analyser-based and grating-based imaging which require multiple sample exposures to extract the phase and dark-field signals. Another advantage of single-grid imaging is that it has a relatively simple setup compared to analyser-based and grating-based imaging techniques since it only requires one optical element - a grid - and no calibration or alignment is required prior to the data acquisition process. In addition, we are able to tune the sensitivity of the system to the dark-field signal, decreasing the pixel size and/or increasing the propagation distance to increase sensitivity. 

We have chosen an approach that uses spatial-mapping/explicit-tracking by cross-correlation, as used for phase retrieval by Morgan \textit{et al.} \cite{morgan2011_grid}, instead of a Fourier approach \cite{takeda1982}, since the Fourier analysis method can fail when the grid frequency overlaps with the sample feature frequency \cite{morgan2011_grid}. The spatial mapping method can also provide better spatial resolution than the Fourier analysis (see \cite[Figure~4]{morgan2013}) since we compare the windows pixel-by-pixel instead of only focusing on a small area in the Fourier space. The spatial mapping method can also be applied on images taken with other reference patterns, such as the sandpaper as shown in Section \ref{sec:speckle}. 

However, the single-grid imaging technique does have a few disadvantages. One of the major disadvantages is that this technique requires a relatively small pixel size to capture both phase and dark-field signals. Generally, the pixel size has to be smaller than the blurring width to obtain high quality dark-field images. Thus, this technique is best-suited to small samples (e.g. millimetre-to-centimetre scale) due to the requirement for a small pixel size. Although it is technically possible to image large samples, which could be achieved by using a detector with a large area but small pixel size, it may not be feasible due to the cost, availability or radiation dose.    

\subsection{Future directions}
In addition to the directions mentioned above, there are a few other possible directions for future investigation. One subsequent investigation would be to explore how the dark-field signal and the scattering angle change with the microstructure size. This could be done by imaging additional samples over a range of microstructure sizes. We could also investigate the dependence of the dark-field signal on the microstructure material and/or the type of medium surrounding the microstructures. 

Other possible future directions include exploring the relationship between the dark-field signal with x-ray energy by performing the experiment with an energy-sensitive detector \cite{taphorn2020grating, sellerer2020}. Finally, an investigation could look at quantifying the edge scattering effects.   

\section{Conclusion}
The x-ray dark-field signal that originates from the SAXS or USAXS of unresolved sample features can provide complementary information on sample properties that is inaccessible from the absorption- and phase-contrast images. A quantitative dark-field signal can be useful in various fields including medical diagnosis \cite{kitchen2020emphysema}, material science \cite{valsecchi2020} and airport security \cite{miller2013phase}. 

In this manuscript, we proposed a new method to extract and quantify the dark-field signal from a single-grid imaging set-up, utilising a mathematical model and a numerical analysis algorithm. The dark-field signal was quantified by curve-fitting and comparing the cross-correlation results between the grid-only image with itself and between the grid-only image and sample-grid image. The quantitative dark-field signal was then converted into the effective scattering angle and then related to the number of microstructures, \(N\), that the beam passes through. The scattering angle was found to be directly proportional to \(^{2.19}\sqrt{N}\), which was roughly consistent with the model proposed by von Nardroff which suggests that the scattering angle is proportional to \(\sqrt{N}\), and matched closely with analyser-crystal-based measurements obtained by Kitchen \textit{et al.} \cite{kitchen2020emphysema}. We showed that the algorithm is able to detect edge scattering and is also compatible with speckle-based imaging \cite{morgan2012, berujon2012_sb, zdora2018state}. 

The feasibility of quantifying the dark-field signal using a single sample exposure was investigated in this manuscript. It was observed that the propagation distance has played an important role in this. We hope that this easily-implemented single-grid dark-field imaging approach can be of use in quantitatively imaging mm-cm-sized samples, particularly in the case of time-sequence imaging or dose-sensitive applications. 

The analysis could be improved by incorporating the propagation-based phase contrast effects into the model through the X-ray Fokker-Planck equation. Further studies can also be done to find optimal propagation distance for single sample exposure imaging of a given sample, to extend into two-dimensions to extract a directional dark-field signal and to investigate the dark-field signal dependence on sample microstructure size, material and x-ray energy.

\section*{Acknowledgements}
We acknowledge useful discussions with David Paganin and recommended references from Marcus Kitchen. We thank Celebrity Groenendijk for sharing the data she had previously published\cite{groenendijk2020material}, which was analysed in Fig. \ref{fig:gold_pmma_res} here and captured at BL20XU at SPring-8 under application 2019A1151. We also thank Sebastien Berujon, Eric Ziegler and Emmanuel Brun for sharing the speckle-based x-ray dataset analysed in Fig. \ref{fig:SBI_raw}, originally published in \cite{berujon2016x} and also in \cite{pavlov2021directional}. The key images in this paper were captured at the Imaging and Medical Beamline at the Australian Synchrotron, part of ANSTO. 

\section*{Funding}
Dr Morgan was supported by FT180100374 from the Australian Research Council.

\section*{Disclosures}
The authors declare no conflicts of interest. 

\section*{Data availability} 
Data underlying the results presented in this paper are not publicly available at this time but may be obtained from the authors upon reasonable request.

\section*{} 
See Supplement 1 for supporting content.

\bibliography{ref_list}

%%%%%%%%%%%%%%%%%%%%%%% References %%%%%%%%%%%%%%%%%%%%%%%%%

%%%%%%%%%% If using BibTeX:

%%%%%%%%%% If preparing manually:
% \newcommand{\enquote}[1]{``#1''}

% \bibitem{Zhang:14}
% Y.~Zhang, S.~Qiao, L.~Sun, Q.~W. Shi, W.~Huang, L.~Li, and Z.~Yang,
%   \enquote{Photoinduced active terahertz metamaterials with nanostructured
%   vanadium dioxide film deposited by sol-gel method,}
%   {\protect\JournalTitle{Optics Express}} \textbf{22}, 11070--11078 (2014).

% {Optical Society}, \enquote{{OSA Publishing},}
%   \url{http://www.osapublishing.org}.

% \bibitem{FORSTER2007}
% P.~Forster, V.~Ramaswamy, P.~Artaxo, T.~Bernsten, R.~Betts, D.~Fahey,
%   J.~Haywood, J.~Lean, D.~Lowe, G.~Myhre, J.~Nganga, R.~Prinn, G.~Raga,
%   M.~Schulz, and R.~V. Dorland, \enquote{Changes in atmospheric consituents and
%   in radiative forcing,} in \enquote{Climate Change 2007: The Physical Science
%   Basis. Contribution of Working Group 1 to the Fourth assesment report of
%   Intergovernmental Panel on Climate Change,}  S.~Solomon, D.~Qin, M.~Manning,
%   Z.~Chen, M.~Marquis, K.~B. Averyt, M.~Tignor, and H.~L. Miler, eds.
%   (Cambridge University Press, 2007).

% \end{thebibliography}

\end{document}

\maketitle

\section{Fitting the cross-correlation result}

\begin{figure}[htbp!]
\centering
\fbox{\includegraphics[width=\linewidth]{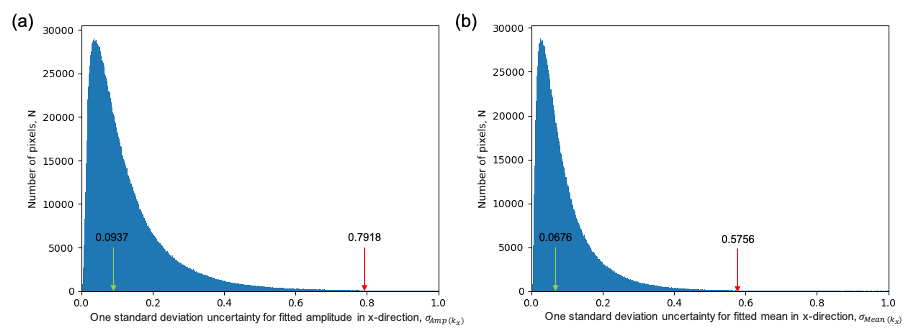}}
\caption{Histogram indicating the confidence of cross-correlation fits, measured as one standard deviation uncertainty for the fitted (a) amplitude, \(\sigma_{Amp(k_{x})}\) (Eqn. 5 in the main text) and (b) mean, \(\sigma_{Mean(k_{x})}\) (Eqn. 6) in the $x$-direction from the cross-correlation analysis between sample-grid and grid-only images taken at propagation distance of 0.2m. The red arrows labelled the \(\sigma_{Amp(k_{x})}\) and \(\sigma_{Mean(k_{x})}\) values obtained from Fig. \ref{fig:sg_gg_cc_res_1300} (c), which is an example where the profiles are not well-described by the fit. Only 0.7\% of the pixels in the image had a worse fit for \(Amp(k_{x})\) and \(Mean(k_{x})\) than the fit shown in Fig. \ref{fig:sg_gg_cc_res_1300}. The green arrows show the median values of the histogram in (a) and (b), which indicates how well the cross-correlation can be fitted with the model on average for a given pixel. The cross-correlation result and the fitted profiles in $x$- and $y$-direction for a pixel with this median confidence fit is shown in Fig. \ref{fig:sg_gg_cc_res_873}. Note that we were only plotting \(\sigma_{Amp(k_{x})}\) and \(\sigma_{Mean(k_{x})}\) values within the range of 0 to 1 for better visualisation purposes and the number of pixels with \(\sigma\) values greater than 1 were negligible.} 
\label{fig:perr_res_27}
\end{figure}

\begin{figure}[htbp!]
\centering
\fbox{\includegraphics[width=\linewidth]{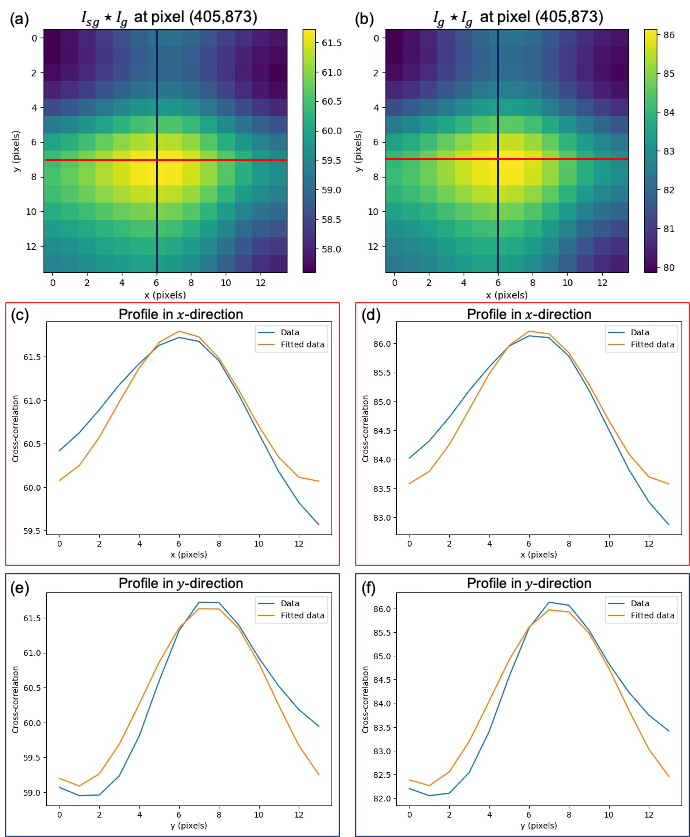}}
\caption{A typical example of the cross-correlation result between (a) the sample-grid image window and grid-only image window, and (b) the grid-only image window with itself, taken at a propagation distance of 0.2m for pixel (405,873) where microsphere sample is present. The maximum score is located at (a) position (7,6) (b) position (7,6). Profiles showing how well the fitted cosine functions (with fixed period and phase shift) agree with the data profiles in (c) \& (d) $x$- and (e) \& (f) $y$-directions. The fitted amplitude and mean values of each panel are (c) 0.8670 and 60.9294 (d) 1.3274 and 84.9046 (e) 1.2878 and 60.3754 (f) 1.8739 and 84.1306 respectively. The dark-field signal, \(DF\) of this pixel (calculated following the method explained in Section 4.1 in the main text) is 0.9379. The decrease in the mean and amplitude value of the plots from (c) and (e) compared to (d) and (f) can be attributed to the attenuation and the scattering effect introduced by the sample on the intensity oscillation. The amplitude and median obtained from (c) has an uncertainty value of 0.0937 and 0.0676 respectively, which corresponds to the median (green arrows) of the histogram in Fig. \ref{fig:perr_res_27}. This figure demonstrates how well the profiles in both directions are fitted on average.}
\label{fig:sg_gg_cc_res_873}
\end{figure}

\begin{figure}[htbp!]
\centering
\fbox{\includegraphics[width=\linewidth]{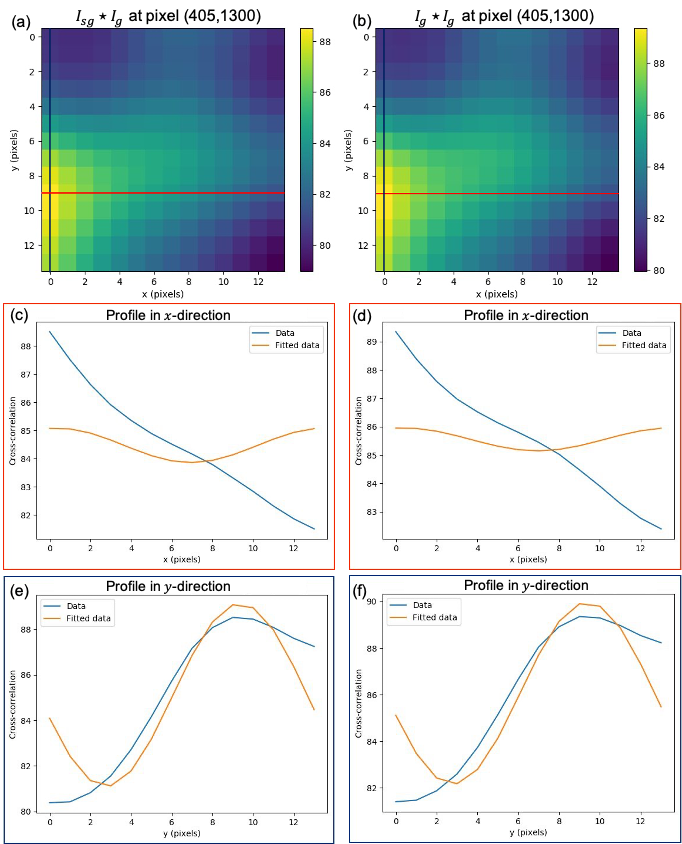}}
\caption{An example of the cross-correlation result between (a) the sample-grid image window and grid-only image window, and (b) the grid-only image window with itself, taken at propagation distance 0.2m for pixel (405,1300) where sample is not present. The maximum score is located at (a) position (9,0) (b) position (9,0). Profiles showing how well the fitted cosine functions (with fixed period and phase shift) agree with the data profiles in (c) \& (d) $x$- and (e) \& (f) $y$-directions. The fitted amplitude and mean values of each panel are (c) 0.6115 and 84.4767 (d) 0.4043 and 85.5473 (e) 4.0146 and 85.1097 (f) 3.9005 and 86.0565 respectively. The dark-field signal, \(DF\) of this pixel is 1.0445. The mean value of the profiles in (c) and (e) remains the same compared to (d) and (f), indicating there is minimum attenuation as predicted by the model since no sample is present. Although the profiles, especially in $x$-direction, are not perfect cosine functions, due to an obstruction on the detector, the algorithm managed to consistently fit a reasonable cosine function for each of them. This is an example of the case where the model does not fit the profile closely due to the deviation of the intensity oscillation from the sine function, but according to Fig. \ref{fig:perr_res_27}, this only accounts for around 1\% of the pixels in the image.}  
\label{fig:sg_gg_cc_res_1300}
\end{figure}

Additional measures are provided here to quantify the performance of our algorithm (Fig. \ref{fig:perr_res_27}) in fitting the cross-correlation profiles (Fig. \ref{fig:sg_gg_cc_res_873} and \ref{fig:sg_gg_cc_res_1300}) to extract the dark-field signal. Even for profiles when the confidence of the fit is low (see the red arrow in Fig. \ref{fig:perr_res_27}, corresponding to the fit in \ref{fig:sg_gg_cc_res_1300}), our algorithm is able to return reasonable fits and is robust to intensity oscillations that are not well-described by a sinusoid.   

\section{Relating the scattering angle to sample thickness}
\begin{figure}[htbp!]
\centering
\fbox{\includegraphics[width=\linewidth]{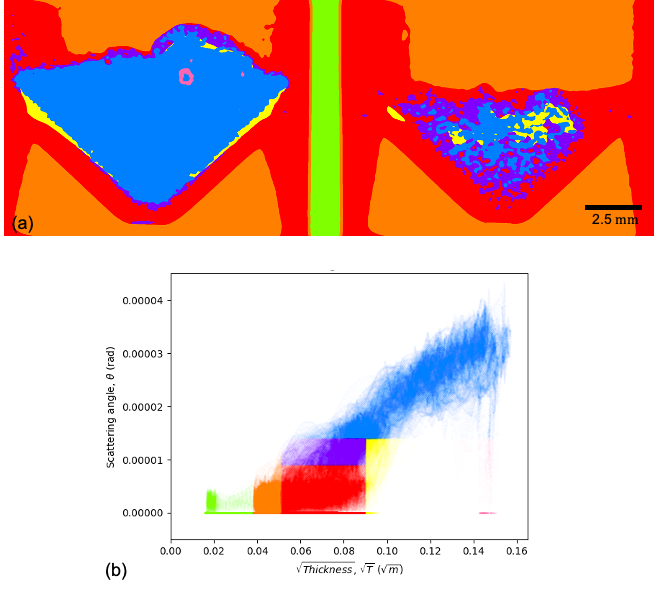}}
\caption{(a) Colour image showing where each data point in (b) the linearized plot of \(\theta\) against \(\sqrt{T}\) is located within the image, where the thickness, \(T\) used in (b) was the thickness recovered from the transmission image using the TIE-based phase retrieval (Paganin et al., 2002), without any correction for the thickness of simulated sample tube. Various features of the sample were segmented in different colours based on the segmentation of the plot in (b). The majority of the microspheres were labelled in blue (with a strong dark-field signal), the edges of the sample tube with greater thickness were labelled in red (predominantly attenuation effects), the unoccupied parts of the sample tube were labelled in orange (with little absorption) and the area where no sample was present was labelled in green. Note that the size of the data points in the plot has been increased relative to the figures in the main paper for better visualisation.}  
\label{fig:theta_color_full_res}
\end{figure}

\begin{figure}[htbp]
\centering
\fbox{\includegraphics[width=\linewidth]{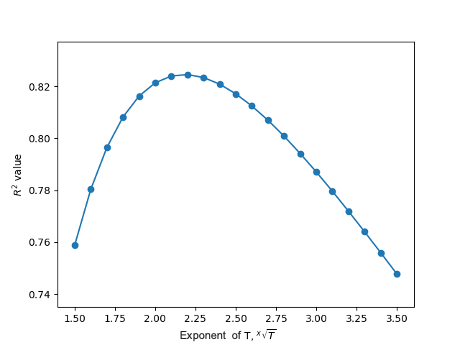}}
\caption{The coefficient of determination (\(R^{2}\)) value of the line of best fit in the linearized plot of \(\theta\) against \(^{p}\sqrt{T}\) changes, for different values of \(p\). The \(R^{2}\) value reaches the peak slightly beyond \(x = 2.0\), which is in agreement with the result presented in Fig. 5 of the main text.}  
\label{fig:r2_plot}
\end{figure}

Here we provide a colour image and a colour plot (Fig. \ref{fig:theta_color_full_res}) to show the complementarity of the attenuation signal (thickness) and dark-field signal. We also show that the relationship between the scattering angle and thickness is best modelled by \(\theta =  ^{2.19}\sqrt{T}\), which gives the maximum coefficient of determination (\(R^{2}\)) value (Fig. \ref{fig:r2_plot}). 

% Bibliography
%\bibliography{sample}